# From GenderMag to InclusiveMag: An Inclusive Design Meta-Method

Christopher Mendez1, Lara Letaw1, Margaret Burnett1, Simone Stumpf2, Anita Sarma1, Claudia Hilderbrand1

1-Oregon State University. Corvallis, OR, United States. 2-City, University of London. London, United Kingdom.

*Abstract*— How can software practitioners assess whether their software supports diverse users? Although there are empirical processes that can be used to find "inclusivity bugs" piecemeal, what is often needed is a systematic inspection method to assess software's support for diverse populations. To help fill this gap, this paper introduces InclusiveMag, a generalization of GenderMag that can be used to generate systematic inclusiveness methods for a particular dimension of diversity. We then present a multi-case study covering eight diversity dimensions, of eight teams' experiences applying InclusiveMag to eight under-served populations and their "mainstream" counterparts.

## I. Introduction

Designing software so that it works for diverse populations matters—to software companies' profitability, to equity in the workplace and at home, and to anyone in a situation that changes the way they think, such as when under deadline pressure. Unfortunately, most software does not support diversity well.

For example, approximately half of all software users are women, but there is increasing evidence that they can be disadvantaged by the way that software is designed [6, 38, 39, 45]. There is also work to include the "missing 20%" of disabled users that are frequently under-served by today's software [17]. There are many other diverse population segments that are currently not included in the design considerations for software. This can lead to lower profitability of the software product, deficient access and accessibility to software by the under-served populations, and reduced usability for all [14, 17].

Inclusive design aims to address this problem by considering diverse users throughout the software design process [11]. There are many ways to bring diverse users into the conversation when designing software. For example, in co-design diverse users can be invited into design sessions to directly collaborate with software designers and one another in a small group setting [7, 29]. Another example is user testing, which can give diverse users an opportunity to provide input about an existing software design, leading to a more inclusive design [31].

However, working with diverse users directly is costly, both in terms of money and time, so methods that do not directly require users to be present are also needed. Toward that end, there has been a move to develop inclusive design guidelines and analytic methods but, except for a few well-researched user groups [42], this work is still in its infancy. Moreover, few of these methods are usable by *software practitioners* in their every-day practice, but instead rely on experts to apply these guidelines and analytic methods.

In this paper, we introduce InclusiveMag (Inclusiveness Magnifier), a (meta-)method to generate inclusiveness methods. We built InclusiveMag inductively, by generalizing upon the principles and processes used to create GenderMag [10]. Our inductive process is similar to one defined by Sjøberg et al. [37] on how theories (and methods) can be inductively defined from concrete practice to more generalized forms.

The InclusiveMag method allows *inclusivity researchers* to set up a systematic inclusiveness inspection method, for *software practitioners* to then apply to their own software to systematically evaluate how it supports (or doesn't) diverse populations. The contributions of this paper are:

- The InclusiveMag methodology, a systematic meta-method for *inclusivity researchers* to generate inclusive design methods for under-served software users;
- A methodology for *software practitioners* to *use* these generated methods to evaluate and re-design their software to increase its inclusivity;
- An early multi-case study of eight teams generating and using the InclusiveMag methodology.

## II. Background

Although InclusiveMag has not been described in the literature, we have been developing it for several years, and have used it to generate GenderMag.

GenderMag, short for "Gender Inclusiveness Magnifier" [10], integrates a specialized cognitive walkthrough (CW) with research-based personas that capture individual differences in how people problem solve and use software features—differences that statistically cluster by gender. GenderMag has been used to detect gender biases in several commercial and open source software products (e.g., [8, 9, 13, 18, 24, 35]).

The GenderMag method rests on five problem-solving facets, which it brings to life with three multi-personas—"Abi", "Pat(ricia)/Pat(rick)", and "Tim". They are multi-personas in that their backgrounds, photos, job titles, etc., are customizable. The facets, however, are fixed. Abi's facet values (Figure 1) are more frequently seen in women than other genders, and Tim's facet values are more frequently seen in men than other genders. The Pats' (identical) facet values emphasize that differences relevant to inclusiveness lie not in a person's gender identity, but in the facet values themselves [19]. GenderMag's personas and facets are integrated into a specialized CW [43].

5/7/19 1:59 PM

## III. THE INCLUSIVEMAG METHOD

InclusiveMag is a (meta-)method to enable inclusivity researchers to generate new inclusive design methods. The methods they generate are then intended for use by software practitioners to evaluate the software they are producing, with the goal of making the software more inclusive to an under-served population, while simultaneously making the software more usable to a mainstream population. As Figure 2 shows, InclusiveMag has three steps—(1) Scope, (2) Derive, and (3) Apply. Inclusivity researchers perform Steps 1 and 2, and software practitioners perform Step 3.

### A. Step 1: Inclusivity Researchers Set the Scope

In Step 1, inclusivity researchers scope the inclusiveness method. They select a software type, select a diversity dimension, and perform research on what might affect how populations along the diversity dimension use the software type. The components of this step are iterative and often intertwined. Step 1 results in a set of facet categories (termed "facets" in this paper), which are relevant to both the under-served and mainstream populations, and facet values, which differ between the under-served and mainstream populations. The facets form the core of the InclusiveMag-generated method.

Step 1's research component is labor-intensive, but the resulting facets depend on its quality. The goal is to produce well-established facets in which individual differences (i.e., the facet values) tend to cluster into the under-served population differently than from the mainstream population, and that are relevant to the chosen type of software. It may include a systematic literature review [21], interviews with experts in the software types and members of the under-served population, lab or field studies, etc. For example, the GenderMag research component included reading theories and empirical work in other disciplines to understand gender differences in cognitive styles and attitudes affecting cognition [4], such as in information processing theory [2, 26, 27, 30, 33] and self-efficacy theory [3, 8, 20, 32, 36]. They also did empirical studies (e.g., [3, 5]).

The output of this step is a "small enough" number of facets to keep the method feasible for use by software practitioners. GenderMag, for example, has five facets selected from the larger set of individual difference research results using three criteria [10]. First, (1) the facet needed to have direct implications for software usage. (2) Second, the facet and/or facets' ties with software usage needed to be backed by extensive prior research. (3) Third, the facets needed to be usable by ordinary software developers or user experience (UX) practitioners who had no prior background in gender research or in psychology [10].

### B. Step 2: Inclusivity Researchers Derive the Method

In Step 2, inclusivity researchers use the facets produced in Step 1 to derive customizable personas and an analytic process specialized to their selected diversity dimension. Step 2 begins with projecting (flattening) the values of each facet (category) onto a linear scale for that facet. These scales provide the positioning for the facet values: one value at each "endpoint" of each facet, and one somewhere within, to make clear that the facet values are on a continuum, not binary (yes/no) values. For each facet, the inclusivity researchers assign to the under-served persona facet values that represent the endpoint of the under-served population, and to the mainstream persona the opposite end-point, selecting endpoints that are reasonably common among those populations, not extreme outliers.

The facet values of the middle persona depend on what the data "tell" the inclusivity researcher to do. Sometimes there are interesting points between the two endpoints. For example, GenderMag learning styles had three discrete styles observed: learning by process, learning by tinkering, and learning by mindful tinkering. There being a third unique or interesting point between the endpoints is not always the case, so sometimes the middle persona is assigned one of the endpoints.

For example, consider GenderMag's risk facet as flattened onto a linear scale. Abi's facet value (risk averse) is at one endpoint, Tim's facet value (risk tolerant) is at the other endpoint and Pat (moderately risk averse) is in the middle. As Figure 3 shows, all of these facet values are fairly common among the population of users shown. Table I shows the assignments of all five GenderMag facets' values.

Fig. 1. Abi's background, age, job, ethnicity, pictures, etc., are customizable, but her thinking is defined by the facets (red roundtangles).

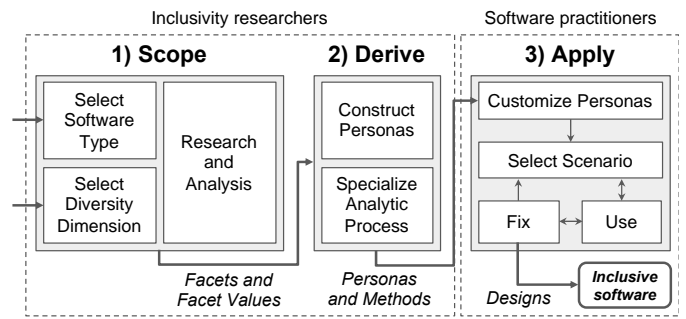

Fig. 2. The InclusiveMag process has three steps, each of which has multiple components. Inclusivity researchers perform Steps 1 and 2, and software practitioners perform Step 3.

The inclusivity researcher then embeds the facets in the different personas, but leaves most of the background section customizable (e.g., Fig. 1) to allow software practitioners to customize the persona in Step 3 to fit their target demographics. For example, in GenderMag, personas' ages, education, job title, familiarity with particular technologies, ethnicity, etc., are customizable, but not the facet values.

For specializing the analytic process, GenderMag specialized a CW, and their procedure generalizes, so we describe it here. (We briefly consider other analytic processes in later sections.)

To specialize a CW, an inclusivity researcher can point explicitly to the selected persona and to relevant facets for each question. For example, as Figure 4 shows, GenderMag researchers specialized in three ways to help software practitioners maintain engagement with the persona [19, 25]. First, the form refers to the persona by name in the questions (Figure 4 (A)). Second, it provides example text to encourage practitioners to express goals/scenarios from the persona's perspective (Figure 4 (B)).

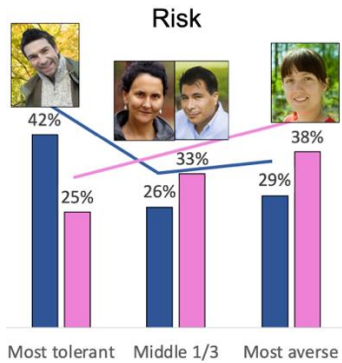

Fig 3. A population of users' self-reported attitude toward risk in technology. Tim represents users on the risk tolerant side of the data, Abi represents users on the risk-averse side, and Pat represents those in the middle.

TABLE I. A SUMMARY OF THE FACET VALUES FOR EACH PERSONA.

| Facet (category) | Abi facet value (Fig. 1) | Pat facet value | Tim facet value |
|---|---|---|---|
| **Motivations** for using technology | Wants what the technology can accomplish. | Wants what the technology can accomplish. | Technology is a source of fun. |
| **Computer Self-Efficacy (confidence)** in using unfamiliar technology | Low compared to peer group. | Medium. | High compared to peer group. |
| **Attitude towards Risk** when using technology | Risk-averse. | Risk-averse. | Risk-tolerant. |
| **Information Processing Styles** for gathering information to solve problems | Comprehensive. | Comprehensive. | Selective. |
| **Learning Styles** for learning new technology | Process-oriented learner. | Learns by tinkering; tinkers reflectively. | Learns by tinkering (sometimes to excess). |

Third, it scaffolds "Why/which" responses with a list of the personas' facets (Figure 4 (C)).

An InclusiveMag CW *itself* needs to be inclusive—collecting a *union* of evaluations, not arguing toward a consensus. To help make this explicit in GenderMag, the forms include a "maybe" option (Figure 4, just below Box "A") to encourage everyone to voice their views along with their explanations of why. Although a potential concern could have been that including all views would encourage false positives (including issues that do not actually arise) GenderMag's empirical false positive rate has been very low, ranging from 0%-4% [10, 41].

### C. Step 3: Software Practitioners Apply the Method

The outcome of Step 2 is a generated method specified to the facets selected in Step 1. In Step 3, a team of one or more software practitioners applies it to their software.

Software practitioners begin Step 3 by customizing the persona(s) they want to use to the appropriate background/demographics/skills for the software they will evaluate (recall Fig. 1). The skills, experience, and education/training dictate what a persona would reasonably be expected to already know and expect to accomplish in the new software features if they haven't used them before.

For example, if software practitioners in Portugal wanted to evaluate a new word processing application using GenderMag, they might make Abi a 35-year old Portuguese novelist who lives in Lisbon and has a degree in creative writing, with experience using other word processing applications. They might also decide to give Tim and Pat the same background as Abi, since the facets already provide the key differences among the personas. Alternatively, they might decide to vary Tim's and Pat's more, perhaps making Tim a 50-year old secretary without a college degree living in Madrid whose writing will be in Spanish, and Pat a 20-year old study-abroad college student from the US living in Lisbon, whose writing might switch between English and Portuguese.

The software team chooses one of the personas they just customized. (One persona is used at a time.) They then choose a scenario to analyze for their software, from the perspective of that persona. For example, a software team using GenderMag might choose Abi for their first session [9]. In the word processing example, a scenario might be "Abi wants to edit Chapter 2's story line to include foreshadowing of an upcoming kidnapping plot. She has already typed in Chapter 2, but hasn't used many of the application's editing features before." Using the persona and the scenario, the team then performs the analysis, producing a list of specific issues that some users like the persona could encounter.

Fig 4. GenderMag's specialization of a CW form (see text).

The session's output is a list of issues to fix. Some of these issues found will be general usability issues (e.g., the font is too small), whereas others will be inclusiveness issues (e.g., risk-averse users would struggle with this step). For the inclusiveness issues, inclusive fixes can be driven by the facets that revealed the issue (e.g., risk). In one GenderMag study, generating fixes to a facet's full range of values (e.g., risk averse and risk tolerant users) resulted in the software improving for everyone, and a previous gender gap in using it entirely disappearing [41].

## IV. AN EARLY MULTI-CASE STUDY OF INCLUSIVEMAG

How generalizable is InclusiveMag? Can inclusivity researchers (other than the original inventors) use InclusiveMag to generate methods analogous to GenderMag, for other diversity dimensions? To find out, we conducted a multi-case study of eight teams using InclusiveMag, who derived eight different InclusiveMag-generated methods.

The setting was an Inclusive Design class for Computer Science upperclassmen and graduate students, a population aiming to become the software practitioners at whom the InclusiveMag method aims. About half the students had Human Computer Interaction (HCI) experience, and some also had professional software development experience. Students formed eight teams of 3-4 people each. All teams included someone with research experience.

Each team worked for 10 weeks. Their goals were: (1) to use InclusiveMag to generate (scope and derive) their method for a software type and an under-served population of their choice along an implied diversity dimension, and (2) to apply that method in an effort to make software prototypes that were inclusive to their under-served as well as a mainstream population.

This empirical set-up involved an empirical trade-off. The disadvantage was that the teams had a relatively concrete focus: to generate a method that would help a single software product's inclusiveness. As Section III shows, the cost of building the method is high enough that many inclusivity researchers would be likely to want a reusable method that could be used on many software products, as with the GenderMag method. However, the empirical advantage to this approach was that it included coverage of how teams went about the third InclusiveMag step, *applying* the generated method to a software product. (It also provided an education advantage: a feedback loop that enabled teams to gain insights into how the method they generated would play out in practice when they had to apply it.)

The eight teams selected a variety of under-served populations and software, such as making email more inclusive for older (and younger) adults; self-driving cars that would work for people with dementia and for people without it; and university websites that would work for people with low socioeconomic status and for people with higher socioeconomic statuses. Table II details the 8 teams' populations and application types.

### A. Step 1: The Teams Set the Scope

#### 1) Scoping the Software Type and Population

All eight teams tended toward a narrow scope for their *software* type (see Table II). This contrasts with GenderMag, for which the software type scope is any "problem-solving software". Had the teams extended their work past 10 weeks, they may have found the narrowness of their software type scope limiting. For example, Team ADHD might want to know how their under-represented persona would fare with Team Autism's math learning app—but since Team ADHD created their persona facets with finance management in mind, the team might have to do the entire InclusiveMag process again, rather than re-using the method they had just generated.

In contrast to narrow software type scopes, some teams scoped their *populations* broadly. For example, Team SES chose people with low socio-economic status for their under-served population. This population is very large and diverse, which could have made it difficult for Team SES to choose a set of facets that was both small enough and sufficiently representative of their under-served population. Even so, because they had chosen a narrow *software* type scope (one section of a university website), they focused most of their research pertinent to students using that university's site, such as basic literacy and digital search skills.

TABLE II. THE EIGHT TEAMS PRESENT IN THE MULTI-CASE STUDY, ALONG WITH SOME INFORMATION ON THEIR PROJECTS. TEAM NAMES USED IN THIS PAPER ARE UNDERLINED.

| Populations considered | Diversity dimension | Software type | Facets from research |
|---|---|---|---|
| <u>ADHD</u>, ≠ADHD | Cognitive | Managing finances | Focus, Organization, Impulsivity, Memory, Financial responsibility |
| <u>Autism</u> kids, ≠Autism kids | Cognitive | Math learning | Comprehension ability, Ability to follow instruction, Concentration level |
| <u>Dementia</u>, ≠Dementia | Cognitive | Self-driving car | Motivations, Memory, Problem-solv. & learning ability, Self-sufficiency/independence, Attention |
| Diabetic <u>retinopathy</u>, Good vision | Vision | Chore robot | Physical/visual ability, Technology preferences, Emotional state & well-being, Financial stability & status, Social interactions |
| Low <u>literacy</u>, Med/High literacy | Education | Language learning | Confidence in using tech, Reading skills, Learning style, Motivations/frustrations with tech. Susceptibility/sensitivity to tech requiring reading |
| Low socio-economic status (<u>SES</u>), Med/high SES | Socio-economic status | University's website | Home life, School experience, Psychological health, Career aspirations |
| <u>Older</u> Adults, ≠Older Adults | Age | Email | Tech. comfortable with, Attitude toward tech, Physical difficulties |
| <u>Pre-schoolers</u>, Adults | Age | Media player | Motivations, Approach to learning, Attitude to recovery, Interaction style, Approach to tech. |

Other teams chose a narrow population slice. For example, Team Retinopathy chose, as their under-served population, a visual impairment resulting from diabetic retinopathy (Figure 11). Diabetic retinopathy is a specific disease that affects, at least to some degree, millions of people (about one-third of the estimated 285 million people in the world with diabetes mellitus) [22]. However, the millions with the disease of diabetic retinopathy are but a small fraction of the approximately 1.3 billion people who have some form of vision impairment [44]. Even more people encounter forms of vision impairment situationally, such as when wearing sunglasses [28].

Despite narrowness's detriment to later reusability of the method they would generate, narrowness had some advantages. For example, during their research into their under-served population, Team Retinopathy identified facets specifically applicable to their population—but not necessarily to other vision impairments—such as emotional well-being (Figure 5). Indeed, in Step 3, this facet did impact the team's design of their prototype:

> Team Retinopathy (excerpt from final report, on design decisions due to facet "emotional well-being"): All of these features will help make Suzie less stressed out as she interacts with the prototype.

*2) Researching the Populations and Facets*

To research their populations, especially the under-served members of it, teams gathered data through literature reviews and, in some cases, directly from individuals in their under-served population. For example, Figure 6 shows an excerpt from Team Older's literature-based research about older adults, and Figure 7 shows summary data gathered by Team SES from individuals in their under-served population.

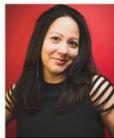

Fig 5. An excerpt from Team Retinopathy's foundation document for Suzie, their under-served persona.

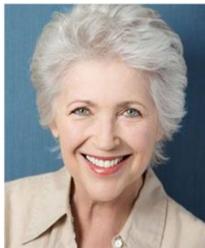

Fig 6. An excerpt from Team Older's persona foundation document with data (highlighted) sourced from literature about their under-served population.

The teams followed a qualitative affinity diagramming process as in [1] to organize their data "factoids" (short facts) into facets (categories) whose values distinguished their mainstream vs. their under-served populations. (In contrast, the GenderMag creators had tended toward quantitative techniques to identify relevant data that clustered by gender, as per Figure 3.)

The facets captured what the teams saw as the most critical attributes of their under-served populations vs. their mainstreamers for their software type scope—thus defining the non-customizable portions of the personas. All eight teams documented the foundations they used to develop the facets via persona foundation documents, which they presented in styles modeled after the GenderMag foundation documents [gendermag.org] or the sample foundation documents in [1].

*3) Which Facets?*

When inclusivity researchers choose how many facets to give personas, they are deciding on behalf of software practitioners, who will need to keep these facets in mind. The GenderMag researchers settled on five facets [10], and the teams loosely patterned their notions on how many facets to choose after that example. Five teams chose five facets, one settled on three facets, and two used four facets.

Team Dementia finessed their five facets by adding 14 subfacets. For example, Figure 8 shows three subfacets within Team Dementia's "Self-sufficiency" facet. An advantage of this level of detail is a rich and informative representation, but a potential disadvantage is the difficulty of keeping 14 subfacets in mind when evaluating a software product. However, Team Dementia's final evaluation explicitly used 11 of their 14 subfacets, and seems to have implicitly used the remaining 3 subfacets.

One reason Team Dementia had so many subfacets may have been because intersectionality was hard for them to avoid. People suffering from dementia are also likely to be older, and both of these situations come with side effects. Team Dementia wanted their facets to be general enough to be reusable but still realistic. Since people with dementia are older, should they also have a motor impairment facet? Since many people with dementia suffer other mental issues as well, such as depression, should depression be a facet?

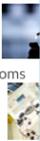

Fig. 7. Excerpts from Team SES qualitative experiences with low-SES people.

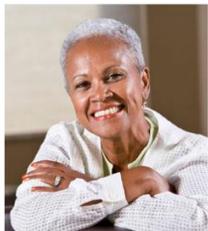

Fig. 8. An excerpt from Team Dementia's foundation document for their under-served persona, showing the multiple subfacets of "Self-sufficiency"

Teams addressed their intersectionality dilemmas in three ways. Some teams, like Team Dementia, incorporated depression into relation to an existing facet value (see the "Living Ability" subfacet in Figure 8). Some teams, when the side effect was not directly associated with the diversity dimension (e.g., an explicit motor impairment), excluded it for generality reasons. Some teams made facets to address physical or mental issues that affect their population, without labeling them with specific disorders. For example, Team Older used the facet "Physical Difficulties" and Team SES used "Emotional Volatility". (We will return to intersectionality in Section V.)

All GenderMag facets are cognition-based, but some of the teams' facets weren't. For example, Team SES had "Home life" and "School experience" (Table II) and Teams Retinopathy and Older included pertinent physical/physiological attributes.

However, Team Older may have gone too far in the direction of concreteness with their "technology she is comfortable with" facet choice (Figure 6). Including specific technology preferences like the ones in the Figure 6 seems likely to give the generated method itself a short 'expiration date'. Such concreteness is common in *personas* for use in a specific *product* line, the traditional use of personas [1]. However, for a *facet* used within a generated *method*, a higher level of abstraction may be called for. For example, "Attitude toward getting the latest technology" might be a more generalizable facet, with the specifics of that technology enumerated only during customization of the background section, which occurs just-in-time when a software team is ready to apply the method to a specific product (Step 3).

*B. Step 2: The Teams Derive Their Method*

Using the results from Step 1, each team then derived two personas from the facets—an under-served persona and a mainstreamer—and selected an analytical process to use with these personas and facets.

Deriving two personas from the facets included deciding upon facet *values* to assign to each persona. This challenged some of the teams, because not all facets reduced well to a linear scale. For example, for Team Autism, the "Nick" (Autistic) persona has difficulty when there are multiple attentional demands, whereas "Jane" (the mainstreamer) becomes bored when there is just one task to concentrate on, and this did not reduce well to "low" vs. "high" concentration abilities. Instead, each persona concentrates best under different circumstances. They settled on making the scale instead be *circumstances* under which each concentrate best (Figure 9).

To choose the (analytic) process they would specialize to "drive" their generated method, all eight teams began with a Studio Analysis process. Scaffolded by a class activity, this process took place twice in class meetings, with about a month between them. With this analytic process, teams set up at tables around the room and a group (here, the members of the other teams) stopped by for informal descriptions (walkthroughs) through the prototype use-cases, with the persona nearby, and provided feedback on problems or opportunities they saw.

In addition to the Studio Analysis process, two teams also specialized another analytic process. Team Literacy specialized a CW during a class meeting (illustrated in Step 3), and Team SES made their facets into heuristics (Figure 10).

As part of specializing the process, one team specialized their prototype. Team Retinopathy used a visual impairment simulator (Figure 11) to show what their prototype would look like from the perspective of someone with diabetic retinopathy. Using an impairment simulator could be a way to specialize any of the analytic processes in Table III.

There are different advantages to highly structured processes like the CW or Heuristic Evaluation (HE), vs. the more informal Studio Analysis sessions (Table III). Structured processes' systematicness produces a thoroughness hard to match in more informal processes. An advantage of the Studio Analysis sessions

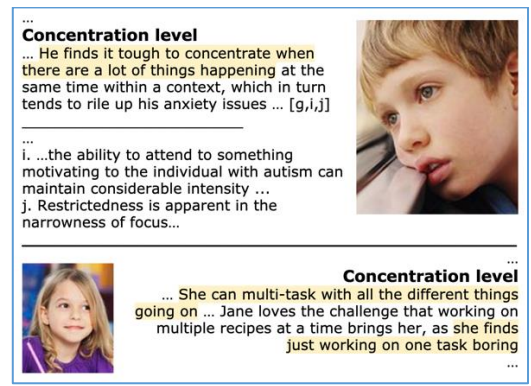

Fig. 9. An excerpt from Team Autism's under-served (top) and mainstream (bottom) persona foundation documents.

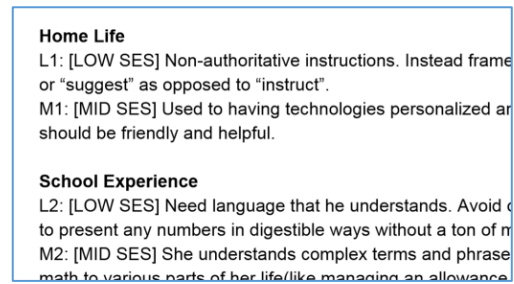

Fig. 10. An excerpt from Team SES's HE process.

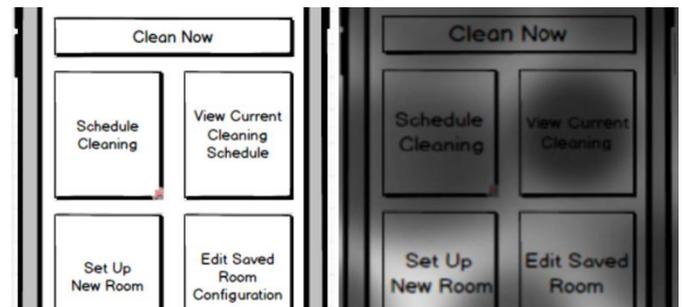

Fig. 11. (Left): An early prototype from Team Retinopathy. (Right): An updated version (larger font) as it could appear to people with diabetic retinopathy, as per the University of Cambridge Impairment Simulator [40].

was that teams got feedback not just on the prototype, but on *all* parts of their method:

**Persona** feedback for Team ADHD: I would avoid using "known" persona pictures to avoid people … overlaying attributes you don't intend for them to have.

**Use Case** feedback for Team Dementia: For use case 2, it seems like making Noah mentally fatigued and tired after work makes your mainstreamer too much like your underserved persona.

**Prototype** feedback for Team Pre-schoolers: children … still easily get lost because of their relatively low comprehension skill. Therefore, if there is a progress bar to indicate their progress toward a specific task, it would be helpful to prevent them from becoming lost.

The above examples suggest that the teams were able to engage with the methods being generated enough to provide feedback on the other teams' emerging methods (facets, personas), methods' application (use cases), and prototypes.

*C. Step 3: The Teams Apply Their Methods*

What kinds of inclusivity issues did the teams find with these methods, and how did they fix them? Here we briefly consider three examples: one from a Studio Analysis-based method (Team Retinopathy), one from a HE-based method (Team SES), and one from a CW-based method (Team Literacy).

From the Studio Analysis process, Team Retinopathy realized how the aesthetics of their robot might actually interfere with the robot's usability or adoption. Their fix, shown in Fig. 12 (left), was based on the following (emphasis added to facet values):

Team Retinopathy: Originally, we … had a claw arm on wheels ... Multiple of our peers pointed out that that design might … negatively impact Suzie's perception of the product, given her **Emotional & Mental Well-Being** facet … <We> changed the design of the robot to SpiderBot … a cute, talking animal-like bot … *[Figure 12]*

Team SES found changes to make based on all eight of their heuristics. For example, two of Team SES's heuristics (Figure 10) came from linguistic facets, which led to them making wording changes (Figure 12, right):

Team SES: Wording: "Your first term may look like" is trying to **be friendly (M1)** and **Non-authoritative (L1)**.

Team Literacy's "Literacy-Mag" CW-based walkthrough occurred during a class meeting, with half the class using GenderMag's Abi persona and the other half using Team Literacy's underserved persona, Dave. Team Literacy used the results of their walkthrough to make changes to their prototype like the one in Fig. 13:

Team Literacy: … our underserved population … <lacks> **confidence in their ability to interact with technological interfaces**, … they often do not know if they … <completed> a task. This screen *[Fig. 13]* offers a confidence boost … and … feedback that they have finished …

This variety of populations, software types, analytic processes used, and fixes generated, provides encouraging evidence of the generality of InclusiveMag, if care is taken with the facets (Step 1), deriving the new methods from them (Step 2) and attending to them (Step 3).

## V. OPEN QUESTIONS

*A. Validating InclusiveMag*

Although the case study data are encouraging, the question of whether InclusiveMag is useful for generating inclusiveness methods that really work is largely open. To consider how the InclusiveMag method might be validated, we follow the lead of Sjøberg et al. [37].

Sjøberg et al.'s recommendations are about validating theories, not validating methods but their validation criteria still provide useful insights into method validation [37]. In Table IV, we consider how to apply these criteria to InclusiveMag, and the available evidence applicable to each.

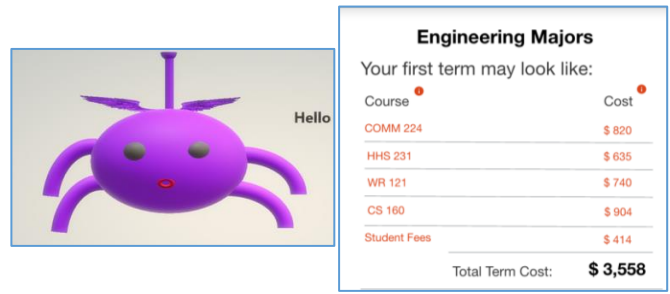

Fig. 12. (Left) An image from Team Retinopathy's final design of spiderbot (Right) Part of Team SES's prototype that underwent a wording change

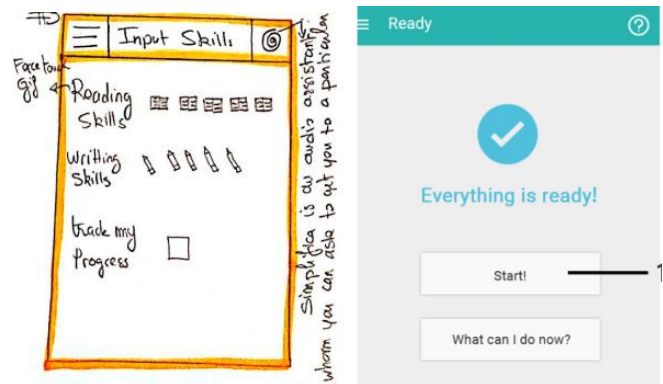

Fig. 13. Screen at the end Team Literacy's use case of customizing the settings. (Left): Before using "Literacy-Mag". (Right): After.

TABLE III. ANALYTIC PROCESSES USED BY CASE STUDY TEAMS

| Teams | Used the analytic process… | Which had the components… | And received feedback on… |
|---|---|---|---|
| (All) | Studio Analysis | Use cases + one or more personas + software prototype | Everything |
| Literacy | Cognitive Walkthrough | Scenario + one persona + software prototype + forms | Prototype |
| SES | Heuristic Evaluation | Scenario + heuristics + software prototype | Prototype |

*B. InclusiveMag in Practice*

One open question is how the facets produced in Step 1 inform Step 2's choice of the analytic process to specialize GenderMag uses strictly cognitive facets, so fits well with including a specialized CW. However, some diversity dimensions like accessibility need physical facets [29], and Team SES had environmental facets (e.g., their "home life" facet). For methods using facets like these, the question in Step 2 of which analytic process to specialize arises. One possibility for some physical attributes may be analyzing with the help of a simulator, as Team Retinopathy did (Figure 11).

Since the facets are the core of InclusiveMag, it seems possible to embed the facets in any analytic process. However, Team SES's attempt to embed their facets in a set of heuristics raises questions as to whether all analytic processes really can support the selected facets well. Team SES's heuristics may have been too low level and overly specific—they focus mostly on language, ignoring other aspects that could also be non-inclusive like icon choices, workflow, etc.

Another question is *how* to actually build a persona into an analytic process other than a CW. Without the persona, the software practitioners lose "theory of mind" benefits (i.e., empathy, or taking another kind of person's perspective), the psychological basis that personas leverage [15].

TABLE IV. APPLYING SJØBERG ET AL.'S EVALUATION CRITERIA TO INCLUSIVEMAG [37]. HERE "ACCURACY" COMBINES PARTS OF SJØBERG ET AL.'S "TESTABILITY" AND "EXPLANATORY POWER" THAT APPLY TO A METHOD

| | "The degree to which..." [37] | Applicability to validating InclusiveMag | Validation evidence to date |
|---|---|---|---|
| Accuracy, Empirical Support | ... empirical refutation is possible. ... supported by empirical studies that confirm its validity. ... predicts all known observations within its scope | Test whether InclusiveMag-generated methods correctly evaluate software's inclusivity. | (1) The only InclusiveMag-generate method that has been tested for validity is GenderMag. Its "true positive" rate at evaluating software inclusiveness has been reported at 75%-100% [10, 41]. (2) For generated versions using CWs: Errors of omission (false negatives) are common in cognitive walkthrough methods, with rates 30%-70%, depending on analysts' expertise [23]. |
| Parsimony | ...<has> a minimum of concepts ... | Investigate whether all steps/components of InclusiveMag are needed | |
| Generality | ...breadth of the scope ... and independent of specific settings | Breadth of scope in (1) InclusiveMag usage, and in (2) InclusiveMag-*generated* methods' usage. | (1) The 8-team case study showed wide breadth of scope for InclusiveMag. (2) The resulting InclusiveMag-generated methods' scopes were explicitly defined (as narrow or broad) by teams generating them. |
| Utility | ...supports the relevant areas of the software industry | Investigate whether software practitioners choose to use the generated methods | |

Finally, could inclusivity researchers leverage personas they already have in InclusiveMag? For example, would they be able to start at Step 2 with their existing persona in hand? We believe that the existing persona might be blendable with the facets, but the facets would need to be thoroughly reconsidered, which may require a repeat in Step 1. Exactly *how* a researcher can decide whether to return to Step 1, and how exactly to go about it in these circumstances is an open question.

*C. InclusiveMag and Intersectionality*

Intersectionality considers specific insights and problems that arise at the *intersections* of two or more different diversity dimensions [34]. Intersectionality is a term originally coined to show how, through only considering race or gender, the experiences of black women were being ignored by anti-discrimination legislation [12]. From this origin, the idea has been adopted by other fields, including HCI [34].

This raises the question of whether it would be possible for InclusiveMag to generate an *intersectional* inclusive design method. One possibility, similar to what we saw teams do in Section IV, is to simply use the scoping process (i.e., Step 1) to define any population of interest (e.g., low-SES women). This possibility may be viable when the under-served population of interest is large, but runs the risk of comparing a smaller of-interest group with "everyone else", which could be problematic (as well as some of the same problems of a narrow population scope seen in Section IV).

A more genuinely intersectional approach seems to require adding more diversity dimensions to InclusiveMag. It remains an open question whether it is possible to expand the number of dimensions, to how many, how to do so, and what the impacts on applying the generated method (Step 3) would be.

## VI. CONCLUSION

In this paper, we have introduced InclusiveMag, a systematic (meta-)method for inclusivity researchers to generate new inclusive methods. These generated methods are then used by software practitioners to evaluate the software they are creating.

In a multi-case study, eight teams used InclusiveMag to generate inclusivity methods along eight diversity dimensions, and then applied their generated methods to their software prototypes. Although the case study is early, it contributes encouraging evidence as to InclusiveMag's generality.

We emphasize that the first two steps of InclusiveMag method are for industrial (or academic) researchers, not for practitioners. However, the case study shows that InclusiveMag may also be useful to professors teaching classes on HCI research methods.

As others begin to use InclusiveMag to generate new methods (Step 1 and 2), the methods they generate will cover more diversity dimensions. These additional methods and dimensions will then enable software practitioners (Step 3) to cover more diversity dimensions—*early* in the lifecycles of the software they create. We believe that enabling this kind of early evaluation of software inclusivity is key to chipping away at software's implicit biases, one inclusiveness issue at a time.

[42] Web Content Accessibility Guidelines (WCAG) 2.0, 2008. [Online]. Available: https://www.w3.org/TR/2008/REC-WCAG20-20081211/. [Accessed: April 20, 2019].

[43] C. Wharton, J. Rieman, C. Lewis, and P. Polson, The cognitive walkthrough method: a practitioner's guide, John Wiley & Sons Inc., 1994.

[44] World Health Organization, Blindness and vision impairment, 2018. [Online]. Available: https://www.who.int/news-room/fact-sheets/detail/blindness-and-visual-impairment. [Accessed: April 23, 2019].

[45] G. Williams, Are you sure your software is gender-neutral? ACM Magazine Interactions, vol. 21, iss. 1, pp. 36–39, January 2014.